\documentstyle[emulateapj,epsfig]{article}

\def\lax    {\ifmmode{_<\atop^{\sim}}\else{${_<\atop^{\sim}}$}\fi}
\def\gax    {\ifmmode{_>\atop^{\sim}}\else{${_>\atop^{\sim}}$}\fi}
\def\kms    {\ifmmode{{\rm ~km~s}^{-1}}\else{~km~s$^{-1}$}\fi}

\def\bk{\lower 6pt\hbox{${\buildrel k\over \sim}$}}
\def\bv{\lower 6pt\hbox{${\buildrel v\over \sim}$}}

\def\blankline  {\vskip10truept}

\begin{document}


\submitted{Astrophys. J. Lett., v. 537, July 10 2000} 
\title{Signatures of Exo-solar Planets in Dust Debris Disks}

\author{Leonid M. Ozernoy\altaffilmark{1,2}, Nick N. Gorkavyi\altaffilmark{2,3},
 John C. Mather\altaffilmark{2} and Tanya A. Taidakova\altaffilmark{4}
}  

\altaffiltext{1}{School of Computational Sciences and Department of 
Physics \& Astronomy, 5C3, George Mason U., Fairfax, VA 22030-4444;
ozernoy@science.gmu.edu}

\altaffiltext{2}{Code 685, Laboratory for 
Astronomy and Solar Physics, NASA/Goddard Space Flight Center, Greenbelt,
MD 20771; \\ozernoy@stars.gsfc.nasa.gov, gorkavyi@stars.gsfc.nasa.gov,
john.c.mather@gsfc.nasa.gov} 

\altaffiltext{3}{NRC/NAS}

\altaffiltext{4}{Computational Consulting Services, College Park, MD 20740;
simeiz@aol.com}

\centerline{\it Received 1999 August 25; accepted 2000 May 31; published} 

\begin{abstract}

We apply our recently elaborated, powerful numerical approach to the
high-resolution modeling of the structure and emission of circumstellar
dust disks, incorporating all
relevant physical processes. Specifically, we examine the resonant
structure of a dusty disk induced by the presence of one planet.
It is shown that the planet, via resonances and gravitational scattering,
produces (1)~an asymmetric resonant dust belt with one or more clumps
intermittent with one or a few off-center  cavities;
and (2) a central cavity void of dust.
These features can serve as
indicators of a planet embedded in the circumstellar dust disk and, moreover,
can be used to determine its major orbital parameters  and even the mass of 
the planet. The results of our study reveal a remarkable
similarity with various types of highly asymmetric circumstellar disks
observed with the James Clerk Maxwell Telescope around Epsilon Eridani and 
Vega. The proposed interpretation of the clumps in those disks as being 
resonant  patterns
is testable -- it predicts the asymmetric design around the star
to revolve, viz., by $1.2^\circ$--$1.6^\circ$ yr$^{-1}$ about Vega and
$0.6^\circ$--$0.8^\circ$ yr$^{-1}$ about $\epsilon$ Eri.

\end{abstract}

\noindent {\it Subject headings}: circmstellar matter -- dust, 
extinction -- planetary systems --\\ 
\hphantom{\it Subject headings}: stars: individual (Vega, $\epsilon$ 
Eridani

\section {INTRODUCTION}

A significant limitation
to unambiguous planet detection in circumstellar disks is the
contaminating thermal emission from dust in the target systems.
We propose to turn this hazard into an advantage for detecting and
characterizing planetary systems.The major  goal of the present
Letter is to explore how the presence of the planet(s) impacts the
disk via particular resonances responsible for the
specific asymmetric features in the dust, and how these features 
would enable us to derive the major
parameters of the yet invisible planetary system.
The major obstacle in providing reliable, high-resolution
numerical simulations -- the particle-number limitation -- is solved by our
recently elaborated  very efficient approach to numerical
modeling of distributions of test particles in an external gravitational
field (Ozernoy, Gorkavyi, \& Taidakova 
1998, 2000; Gorkavyi et al. 2000; Taidakova \& Gorkavyi 1999). 
\vskip 0.2truein
\section{ RESONANCE STRUCTURES IN DUSTY CIRCUMSTELLAR DISKS}

Both the Poynting-Robertson (P-R) and stellar wind drags 
tend to induce dust inflow toward the star. As the dust passes by the
planets in its infall, it interacts
with them by accumulating in the outer planetary resonances, which, as we
demonstrate below,
could serve as an efficient means of planet detection.
The existence of a resonance ring associated with the Earth's orbit was
predicted by Jackson \& Zook (1989); such a ring  was indeed discovered
in the {\it IRAS} (Dermott et al. 1994) and  {\it COBE} (Reach et al. 1995) 
data. 
The resonant design in the  dust depends upon the capture probability,
the particle lifetime in the resonance, the amplitude of libration, and 
the typical eccentricity of resonant particles. 
The total mass of dust contained in a particular resonance is the product of
the probability for the dust particle  being captured into the resonance,
$W_j$; the lifetime spent in the resonance, $\tau_j$ (in a sufficiently dense
environment, this time is due to particle collisions); and the total dust
flow per unit time in the vicinity of the resonance, ${\rm d}M/{\rm d}t$, i.e.
$M_j\approx W_j\tau_j ({\rm d}M/{\rm d}t)$. Having introduced, for
convenience, the mass of `background' dust, $M_b\approx (1-W_j)\tau
({\rm d}M/{\rm d}t)$,
one gets $M_j/M_b\approx W_j\left(1-W_j\right)^{-1}\left(\tau_j/\tau\right)$,
where $\tau$ is the minimal timescale of the P-R drag ($\tau_{PR}$),
gravitational scattering ($\tau_{gs}$), and collisional ($\tau_{coll}$)
characteristic times. Three particular cases are well distinguishable:

{1}: $\tau\approx\tau_{PR}<\tau_{gs},
\tau_{coll}$. Actually, this collisionless case is realized, for example, for 
Neptune in the Solar system. Assuming that $W_j\sim 1-W_j$, one gets $M_j/M_b
\sim \tau_j/\tau_{PR}$. To get a discernable resonant structure,
one needs to have $\tau_j\sim\tau_{PR}$.

{2}:
$\tau\approx\tau_{gs}<\tau_{coll}, \tau_{PR}$. This case is realized 
for a rather massive planet like Jupiter. If collisions are essential 
(as they are for Vega's disk, see Backman 1998), then
$\tau_j\approx \tau_{coll}$ and one gets $M_j/M_b\sim \tau_{coll}/\tau_{gs}
\gax 1$. The resonant structure has a high-contrast if $\tau_{gs}\ll
\tau_{coll}$. 

{3}: $\tau\approx\tau_{coll}<\tau_{gs},
\tau_{PR}$.  In this collisional case one has  $\tau_j\approx\tau_{coll}$
giving $M_j/M_b\sim W_j/(1-W_j)$,
which results in a high-contrast resonant structure whenever $W_j\sim 1$.

It is worth emphasizing that if collisions of resonant particles are 
important (cases 2 and 3), the resonant pattern can nevertheless
exist (cf. Wyatt et al. 1999) as long as a source of particle replenishment 
is available.

\section{MODELLING CIRCUMSTELLAR DUST DISKS}

\centerline{3.1. {\it  The Procedure}}

\blankline 
The simplest planetary system under consideration, consisting of just 
one planet (orbiting a star of mass $M$ at a circular orbit
of radius $a$) possesses only mean motion resonances, without most of
secular resonances (except the Kozai resonances).
We have computed about 300 model disks in order
to explore the effects of  various resonances adopting the following 
parameters: (1) the mass of the planet, $m_{pl}$; 
(2) the particle size coupled with  stellar luminosity and mass,
as described by the parameter $\beta
\approx 0.3~L_\star/(M_\star r)$, which is the ratio of the stellar
light pressure and the gravitational force applied to the dust grain
of density 2 g~${\rm cm}^{-3}$ and radius $r$ (in $\mu$m), with $L_\star$
and $M_\star$  the star's luminosity and mass in solar units;
 and (3) lifetime of dust particles, $\tau$.
Below, we describe some of these models.

\blankline 
\centerline{3.2. {\it Typical Resonant Structures Exemplified by the}} 
\centerline{\it Resonances 1:1, 2:1, 3:1, and 3:2}

\blankline
In the absence of a planet, the surface density of dust between the star
and the pericenter of dust sources is constant, while the number density
of dust varies as $r^{-1}$, $r$ being heliocentric distance (Gorkavyi et al. 
1997). The presence of a planet induces dramatic changes in the dust 
distribution. This
is illustrated in Fig.~1, which shows the geometry of dust distribution in
different mean motion resonances with the planet.

In the resonance 1:1, the orbits responsible for the dust design
 are of three kinds:
(1) a horseshoe-like orbit that embraces the two  
Lagrange points  L4 and L5 (see Fig.~1a); (2)    
an orbit with a small libration around L4 (Fig.~1b, {\it upper feature});
and (3) an orbit with a substantial libration around  L5 (Fig.~1b, {\it 
lower feature}).

The dust structure in the resonance 2:1 looks similarly: the inner orbital
loop precesses relative to L4 and L5 in the same way as the epicycle of
particles in the 1:1 resonance. That loop can form a horse-like arc along
the planet's orbit (Fig.~1c) or can librate 
near either L4 or L5 (Fig.~1d illustrates the case of a large libration).

In the resonance 3:1, the dust particles
librate around the L4 with a smaller amplitude (Fig.~1e);
no horse-like libration has been observed. Finally, Fig.~1f illustrates
the resonance 3:2 with an appreciable libration, which results in a
symmetric four-clump pattern.  
When libration is small, the resonance contains a two-clump pattern.

In all cases mentioned above, the vicinity of the planet is free of
dust. Each resonant dust structure is accompanied by 
other specific caverns in dust distribution  (see Fig.~1).
 
\blankline 
\centerline{3.3. {\it Dependence of Structure on Location of Dust Sources}}
\blankline
The scattered cometary population produces dust whose distribution 
in  orbital elements tends to be rather smooth. A well-pronounced
resonant structure in the dust can only be formed if there is a planet that
massive enough to capture some dust particles into resonances and to
eject the remaining non-resonant dust particles out (Case 2 in Sec.~2).

Dust sources like the Kuiper belt objects (KBOs; hereafter `kuiperoids'), 
of which  a substantial part  is
in resonances, can produce the contrast resonant structures if
(1) the drifting dust particles have a long lifetime and eventually
are captured into resonances (the situation is similar
to the formation of a resonant dust  belt near
the Neptunian orbit in the Solar system) or (2) both $\beta$ and the 
lifetime of dust particles are small enough, so that the particles 
produced by the resonant kuiperoids cannot leave the resonances
and thereby they form resonant structures (Case 3 in Sec.~2). Orbits of 
resonant objects look similar to dust orbits shown in Fig.~1.

\blankline 
\centerline{3.4. {\it Dependence of Structure on Planetary Mass}}. 
\blankline
This dependence
can be revealed from a variety of available runs with different parameters.
A low-mass planet, like Earth or Neptune, forms a resonant ring along the
planetary orbit associated with high-order resonances, like 3:2, 4:3, etc. 
(see, e.g. Fig.~1f). For more massive planets,
the resonances like 2:1, 3:1, etc. become essential, whereas
the innermost resonances overlap (e.g. Wisdom 1980).
Since the libration amplitude decreases as $m_{pl}$ increases, 
clumps similar to those
seen in Fig.1e (and not arcs like those shown in Fig.~1d) would
be typical for $m_{pl}$ as large as $m_J$, where $m_J$ is Jupiter's mass.

\section{A LINK TO OBSERVATIONS: IMAGERY OF CIRCUMSTELLAR DISKS}
Let us consider some observational signatures of planets embedded in disks.
Figures 2a and 2b show the thermal emission from the simulated disks (seen 
face-on) with an embedded planet, and these are to be compared with the 
available observational data on Vega and  $\epsilon$ Eri shown in Figures 2c
and 2d, respectively (the parent stars are seen almost pole-on).

As a hotter star, Vega might result in a more extensive, 
compared to $\epsilon$ Eri, dust production from evaporating comets. 
We adopt $\beta=0.3$ (close to an upper limit of $\beta=0.5$ at which 
the dust particle escapes) and $\tau_{coll}=2\times 10^4$ planet revolutions
consistent with $\tau_{coll}\sim 10^7$ yrs (Backman 1998).  
A planet as massive as $\sim 2~m_J$ 
induces the resonances $n:1$, which results in two dusty clumps on either side
of the planet,
while the planet ejects the remaining non-resonant dust particles out 
(Case 2 in Sec.~2). Thus, even if the distribution of the cometary
population around Vega might be symmetric, ejection of non-resonant
particles by the  planet could make the resonant dust pattern highly
asymmetric (Fig.~2a), which is reminiscent of the observed two asymmetric 
clumps near Vega shown in Fig.~2c (Holland et al. 1998).

In contrast to Vega, $\epsilon$ Eri is a less massive and a colder star 
with $L\approx {1\over 180}~L_{\rm Vega}$. Besides, the typical size 
of dust particles is several times larger than in Vega's disk (Backman 1998),
therefore we adopt $\beta= 0.002$ for the dust in $\epsilon$ Eri's disk.
 Assuming that kuiperoids serve as the main source of
dust in this disk, we find that the dust 
\begin{figure*}[t] 
\centerline{\epsfig{file=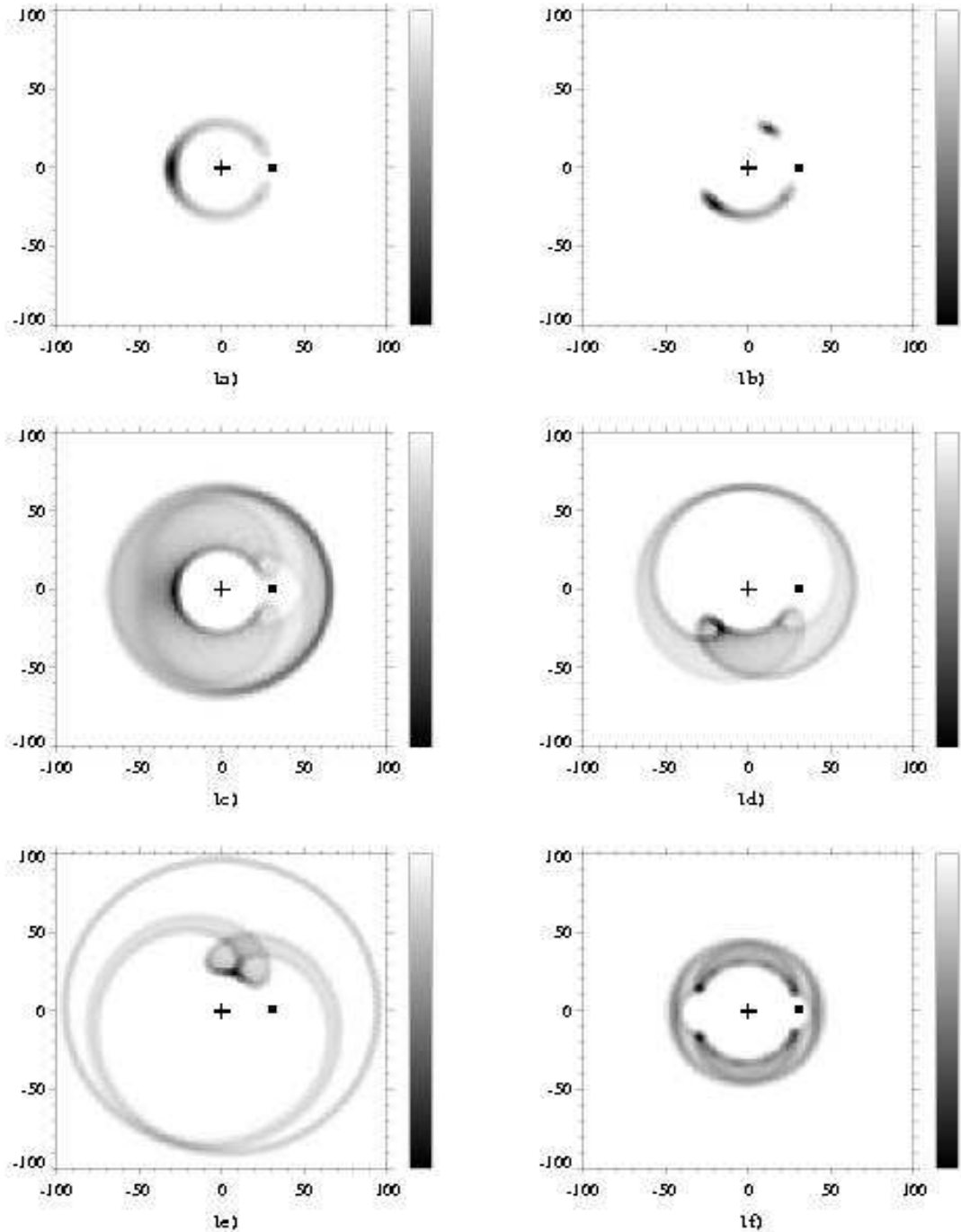,width=5.7in,height=7.5in}}
\figcaption[ogmtfig1.ps]{
Simulated surface density of the circumstellar dust
($\beta=0.065$) captured into particular mean motion resonances with an outer 
planet shown as a square. The star's location is (0,0). In each panel, 
the surface density of the disk is shown on
a linear contrast scale; the area covered is 200x200 (scale is arbitrary
but, for the Solar system, it would be in AU); and the (single)
planet is assumed to be on a circular orbit ($e_{pl}$=0) at $r=30$.
 The  particle lifetime in the resonance,
$\tau_{res}$, was taken to be $10^3$ planet revolutions everywhere:
{\it a} $m=0.3~m_J$, resonance 1:1
(a horseshoe-like orbit around the two Lagrangian points L4 and L5);
{\it b} same as {\it a}, but 
an orbit around L4 with a minimal libration and another, nonoverlapping 
orbit around L5 with a maximum libration;
{\it c} same as {\it a}, but  resonance 2:1; {\it d} $m=0.3~m_J$, 
resonance 2:1 (an orbit around L4 or L5 with a  maximum libration);
{\it e} $m=3~m_J$,  resonance 3:1 (an orbit around the Lagrangian point L4 
with a smaller libration amplitude caused by a larger $m_{pl}$). Note that
at  $m_{pl}$ as small as $m=0.3~m_J$, there is no capture into this 
resonance; 
{\it f}  $m=0.3~m_J$, resonance 3:2.
\label{fig1}}
\end{figure*}
\begin{figure*}[t] 
\vspace{-7.5cm}
\centerline{\epsfig{file=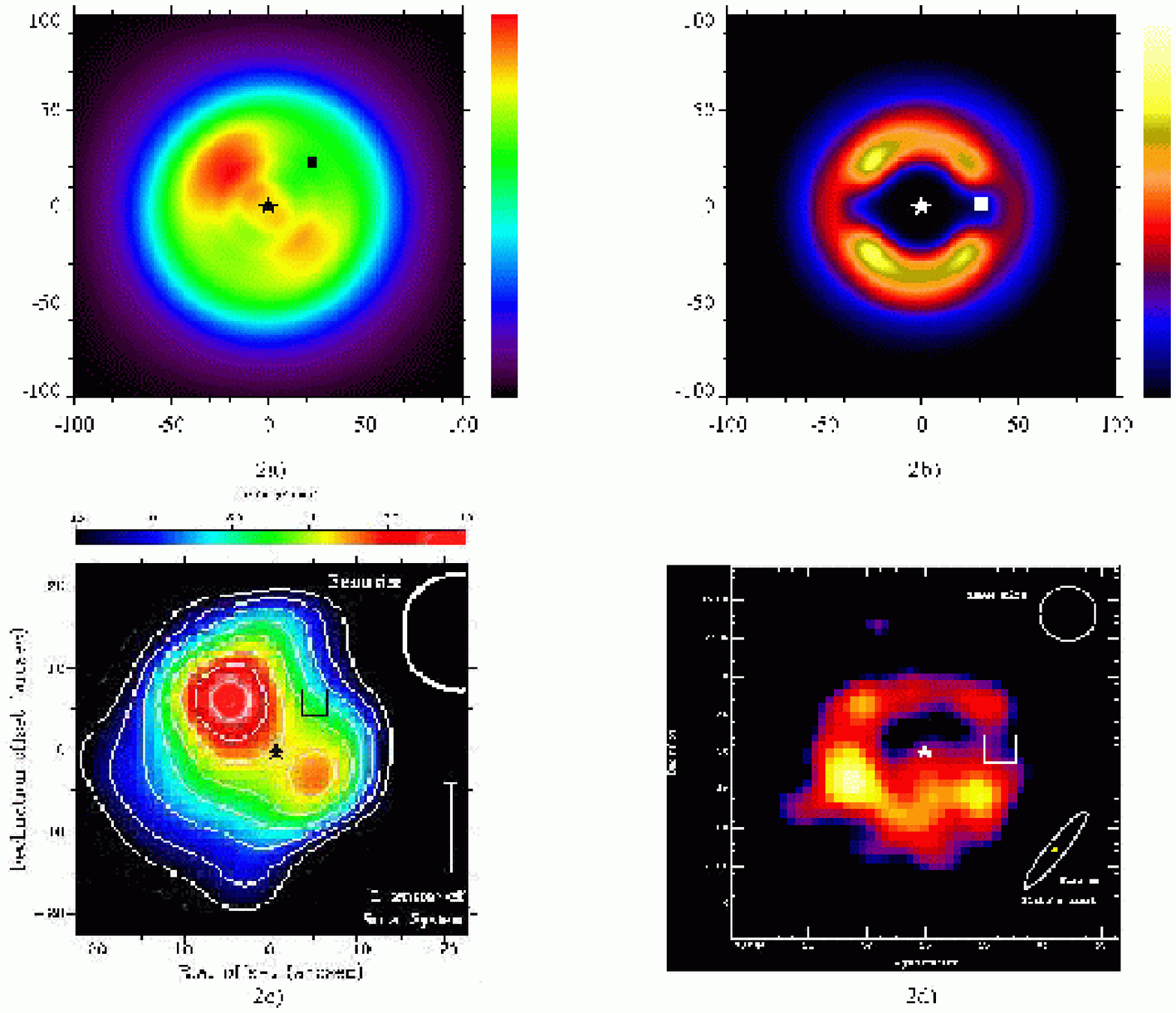,width=7.125in,height=9.19in}}
\figcaption[ogmtfig2.ps]{
Thermal emission ($\lambda = 850~\mu$m) from simulated
circumstellar disks with the resonant structure dominated by a few
resonances vs. observations.
The planet shown in as a  square in {\it a, b}  revolves, like the disk,
counterclockwise. 
In  {\it c, d}, a square indicates the expected planet's location.
 The star's location is (0,0).
To compare our simulations with observations, averaging with an appropriate
beam size is applied to the simulations.
{\it a}
Simulation reminiscent of  the disk of Vega:
$M_\star=2.5~{\rm M}_\odot$;  $m_{pl}=2~m_J$; $\beta=0.3;  \tau=
2\times 10^4$ planet's revolutions; the number of dust sources (comets)
is 486, and the number of particle positions is $3\times 10^{10}$.
\noindent{\it b}.
The simulation reminiscent of the disk of $\epsilon$~Eri:
$M_\star=0.8~{\rm M}_\odot$; $~m_{pl}= 0.2~m_J$; the resonances 2:1 and
3:2 (in equal proportions); $\beta=0.002; \tau= 10^2$ planet's revolutions;
the number of dust sources (kuiperoids)
is 3567, and the number of particle positions is $2\times 10^{9}$.
{\it c}. Circumstellar disk  of Vega (Holland et al. 1998).
{\it d}. Circumstellar disk  of $\epsilon$ Eridani (Greaves et al. 1998).
\label{fig2}}
\end{figure*}
\noindent 
distribution might reflect the distribution of the sources located mostly in 
the resonances such as 2:1 and 3:2. In this case, 
a comparatively small planetary mass of
$\sim 0.2~m_J$  is able to induce, via libration,  two arcs, 
with a pair of clumps at their edges (Fig.~2b). Even if the dust is 
distributed in equal proportions to both resonances, one could easily get an 
asymmetry seen in Fig.~2b by assuming that 10\% of comets in the resonance 
2:1 have a smaller libration amplitude in the lower branch of the pattern.
A similar asymmetric
structure has been revealed in the submillimeter imagery  of the 
$\epsilon$~Eri disk shown in Fig.~2d (Greaves et al. 1998).

Our modeling, which indicates that Vega may have a planet of mass 
$\sim 2~m_J$
at a distance of $50-60$ AU, and $\epsilon$~Eri may have a less massive 
planet of $m\approx 0.2~m_J$ at a similar distance of $55-65$ AU, 
is testable. Each resonant feature is stationary in the
reference frame corotating with the planet, but it is not so for the
observer at Earth. Therefore, if our interpretation of asymmetric
clumps as a 
dynamical resonant structure is correct, the above asymmetric feature 
revolves around the star (probably, counterclockwise) with an angular 
velocity of $1^\circ.2$--$1^\circ.6~{\rm yr}^{-1}$ (Vega) and 
$0^\circ.6$--$0^\circ.8~{\rm yr}^{-1}$ ($\epsilon$~Eri) --
a prediction
that can be tested within several years. If confirmed, the proposed
interpretation  of the structure in Vega- and $\epsilon$ Eri--like
circumstellar disks seen
face-on would make it possible not just to reveal the embedded planet and
determine its semimajor axis, but
also to constrain its other basic parameters, such as the planet's mass
and even to pinpoint the position of the planet (see Figs.~2a and 2b).

\section{CONCLUSIONS}

As we demonstrated above, highly asymmetric structures in the circumstellar 
disk and a central `hole' void of dust can serve as indicators
of at least one planet embedded in the disk. 
The numerical simulations described above lead to the following major
conclusions:

1. The outermost planet in an exo-planetary system
can produce: ({\it a}) an asymmetric resonant dust belt consisting of arc(s) 
and one or more clumps intermittent with one or a few cavities;
and ({\it b}) a central cavity void of dust.

2. The morphology of these belts and cavities (size, asymmetry, number of
clumps, and their pattern) depends upon particular resonances involved and
is eventually a function of the planetary mass, location of dust sources,
and $\beta$.

3. The crucial test of the above picture would be
  the  discovery of revolution of the resonant
asymmetric structure around the star. For circumstellar disks in Vega  and
$\epsilon $Eri  the asymmetric design is expected to revolve,
respectively, by $1^\circ.2$--$1^\circ.6$ and $0^\circ.6$--$0^\circ.8$  
annually.

In sum, our numerical simulations offer a possibility to  reveal the
presence of the planet(s) in a circumstellar dusty disk 
seen pole-on and, under further development, could provide a method for
determination of planetary parameters using the visible morphology of an
outer part of the disk.
The proposed approach seems to be sufficiently powerful to provide 
also the interpretation of disks seen edge-on in scattered
light, such as the $\beta$ Pic disk revealed by STIS optical observations
(Heap et al. 2000). 

\blankline
{\it Acknowledgements}. This work has been supported by NASA Grant NAG5-7065
to George Mason University. N.N.G. has ben supported through NAS/NRC
Associateship Research
program. The authors are thankful to Sally Heap help with programming and
helpful discussions. We thank to W.S.~Holland, J.S.~Greaves and their teams
for permission to reproduce their SCUBA images.
\blankline 

\centerline{REFERENCES}
\blankline 

\def\ref#1  {\noindent \hangindent=24.0pt \hangafter=1 {#1} \par}
\smallskip
\ref{Backman, D.E. 1998, in {\it Exozodiacal Dust Workshop}, eds. 
D.E.~Backman et al. (NASA CP-1998-10155; Moffet Field: NASA/ARC), p.13}
\ref{Dermott, S.F., Jayaraman, S., Xu, Y.L., Gustafson, B.A.S. \&
    Liou, J.C. 1994, Nature 369, 719}
\ref{Gorkavyi, N.N., Ozernoy, L.M., Mather, J.C. \& Taidakova, T. 1997,
 ApJ 488, 268}
\ref{Gorkavyi, N.N., Ozernoy, L.M., Mather, J.C. \& Taidakova, T.,
2000, in {\it NGST} Science and Technology Exposition, eds. E.P.~Smith \& 
K.S.~Long (San Francisco: ASP). ASP Conf. Ser. 207, 462 (astro-ph/9910551)}
\ref{Greaves, J.S. et al. 1998, ApJ 503, L133}
\ref{Heap, S.R. et al. 2000, ApJ, in press}
\ref{Holland, W.S. et al. 1998, Nature,  329, 788}
\ref{Jackson, A.A. \& Zook, H.A. 1989, Nature 337, 629}
\ref{Ozernoy, L.M., Gorkavyi, N.N., \& Taidakova,T. 1998, astro-ph/9812479}
\ref{Ozernoy, L.M., Gorkavyi, N.N., \& Taidakova,T. 2000, Planet. Space
Sci. (in press)}
\ref{Reach, W.T., Franz, B.A., Weiland J.L. et al. 1995, Nature, 374, 521}
\ref{Taidakova, T.A. \& Gorkavyi, N.N. 1999, in
The Dynamics of Small Bodies in the Solar Systems: A Major Key to
Solar Systems Studies, Eds. B.A.~Steves, A.E.~Roy, \& V.G.~Szebehely
(NATO ASI Ser. C, 522; Boston: Kluwer), 393}
\ref{Wisdom, J. 1980, AJ 85, 1122}
\ref{Wyatt, M.C. et al. 1999, ApJ 527, 918}
\end{document}